\documentclass[aps,prd,twocolumn,superscriptaddress,amsmath,amssymb]{revtex4}
\pdfoutput=1
\usepackage{graphicx}
\usepackage{color}
\usepackage{ulem}
\usepackage{float}
\begin{document}
\title{Testing quantum decoherence at DUNE}
\author{J.A. Carpio}
\affiliation{Secci\'on F\'isica, Departamento de Ciencias, Pontificia Universidad Cat\'olica del Per\'u, Apartado 1761, Lima, Per\'u}
\affiliation{Department of Physics, The Pennsylvania State University, University Park, Pennsylvania 16802, USA}
\author{E. Massoni}
\author{A.M. Gago}
\affiliation{Secci\'on F\'isica, Departamento de Ciencias, Pontificia Universidad Cat\'olica del Per\'u, Apartado 1761, Lima, Per\'u}

\begin{abstract}
We address some theoretical issues of the quantum decoherence phenomenon within the neutrino oscillation framework and carry out various 
tests under DUNE simulated experimental environment. On the theoretical side, we provide a general expression for an invariant decoherence 
matrix under a quantum basis rotation. On the simulated experimental side, considering a rotation invariant and a non-invariant 
decoherence matrix, we 
study the impact on the fitting of the standard oscillation parameters, the sensitivity in the mass hierarchy and CP violation, 
combining the neutrino and antineutrino mode and all available neutrino oscillation probabilities channels. Furthermore, a sensitivity for 
the decoherence parameter of order $10^{-24}\text{GeV}$ at 3$\sigma$, is obtained for our best case.
\end{abstract}

\maketitle
\section{INTRODUCTION}
It is well established that neutrino oscillation is induced by non-zero neutrino mass  \cite{Fukuda01,Ahmad02,Fukuda98,Kajita16,Araki05,An12,Adamson14}. 
However, the existance of some, still unrevealed, subdominant mechanism 
is not forbidden. In general, the typical trait of this subleading effects is to the neutrino (oscillation) connection with physics beyond the Standard Model. Within this category there are several theoretical hypothesis such as: neutrino decay~\cite{Berryman15,Frieman88,Raghavan88,Berezhiani92a,Berezhiani92b,Berezhiani93,Barger99,
Beacom02,Joshipura02,Bandyopadhyay03,Ando04,Fogli04,Palomares05,Gonzalez08,Maltoni08,Baerwald12,
Meloni07,Das11,Dorame13,Gomes15,Abrahao15,Picoreti16,
Bustamante17,Gago17,Coloma17,Ascencio-Sosa:2018lbk}, nonstandard neutrino interactions~\cite{Gonzalez99,Bergmann00,Guzzo04,Gago01b,Gago02b,Ohlsson13,Esmaili13}, Lorentz and CPT invariance 
violation~\cite{Colladay:1996iz,Coleman:1998ti,Coleman:1997xq,
Colladay:1998fq,Adamson:2010rn,
AguilarArevalo:2011yi,Li:2014rya}, etc. There is another beyond Standard Model
hypothesis which contemplates an interacting environment due to some effects produced either by strings and branes~\cite{Ellis,Benattistrings} or quantum gravity~\cite{Hawking1}. The result of the interaction between the neutrino system
and the environment is the introduction of decoherence/dissipative parameters into the standard oscillation framework.~\cite{Benatti00,Benatti01,Gago02a,Oliveira10,Oliveira13,Oliveira16,Lisi00,Barenboim05,Farzan08,Bakhti15,
Guzzo16,Gago01a,Morgan06,Fogli07,Oliveira14,Balieiro16,Carpio18,Oliveira18,Coloma18}. This paper is motivated for two interesting observations raised 
in \cite{Carpio18}: the first was the discussion about the invariance of the decoherence matrix when we rotate the neutrino Hamiltonian in matter from the vacuum mass eigenstate basis 
(VMB) to the matter mass eigenstates basis (MMB). The second was the possibility of having a degeneracy in oscillation probabilities between the decoherence parameter 
$\Gamma$ and the CP violation phase $\delta$.  

In this paper we  will present a detailed demonstration of the behaviour of the decoherence matrix under rotations and obtain the general 
shape of a (rotation) invariant decoherence matrix. Furthermore, we will probe the neutrino oscillation probabilities for an invariant and 
non-invariant decoherence matrix under simulated experimental conditions in the context of DUNE, which will have unprecedented sensitivity 
to the identification of the mass hierarchy and the measurement of the CP phase $\delta$~\cite{cdr}.  Considering the two aforementioned 
kinds of decoherence matrix, we will perfom various tests such as: the sensitivity to the 
decoherence parameter, the effects of decoherence in the measurement of oscillation parameters and in the sensitivity to the mass hierarchy and $\delta$. We include
the study of the degeneracy between $\Gamma$ and $\delta$, in agreement with our second motivation point.     

\section{Theoretical considerations}

\subsection{Density matrix formalism}
\label{DensityMatrixFormalism}
The description of the neutrino system weakly interacting with the 
environment using the open quantum system approach is given by: 
\begin{equation}
\frac{d\hat{\rho}(t)}{dt} = -i[H,\hat{\rho}(t)]+D[\hat{\rho}(t)].
\label{LindbladMasterEquation}
\end{equation}
This is the well known Lindblad Master Equation, where 
$\hat{\rho}(t)$ is the neutrino's density matrix and $H$ is the Hamiltonian of the system. The term $D$ encloses the 
dissipative/decoherence effects and is written as:
\begin{equation}
D[\hat{\rho}(t)] = \frac{1}{2}\sum_j \left([\hat{\mathcal{V}}_j,\hat{\rho}(t)\hat{\mathcal{V}}_j^\dag]+[\hat{\mathcal{V}}_j\hat{\rho}(t),\hat{\mathcal{V}}_j^\dag]\right)
\label{lindblad}
\end{equation}
where $\{\hat{\mathcal{V}}_j\}$ is a set of dissipative operators with $j=1,2,\dots,8$ for $3$ neutrino generations. The operators $\{\hat{\mathcal{V}}_j\}$, $\hat{\rho}$ and $H$ can be expanded in terms of the $SU(3)$ Gell-Mann matrices. After, several intermediate manipulations, well explained in~\cite{Carpio18}, we can arrive at the 
solution of the Lindblad Master Equation:  
\begin{equation}
\vec{\rho}(t) = e^{(M_H+M_D)t}\vec{\rho}(0)
\label{exponentialEq}
\end{equation}
where $\vec{\rho}$ is an eight-dimensional column vector consisting of the $\rho_k$, the components of the aforementioned SU(3) expansion,
while the $8 \times 8$ matrices $M_H$ and $M_D$ encode the Hamiltonian and decoherence components of the same expansion, respectively. The dissipative/decoherence matrix $M_D$, that contains all the decoherence parameters, has to be a symmetric, positive-semidefinite
matrix and its entries should satisfy a set of inequalities (see \cite{Hernandez16} for a full list).  
Thus, we can get the oscillation probabilities $P(\nu_\alpha\to\nu_\beta)\equiv P_{\nu_\alpha\nu_\beta}$ 
that can be obtained via inner products
\begin{equation}
P_{\nu_\alpha\nu_\beta} = \frac{1}{3}+\frac{1}{2}\vec{\rho}_{\nu_\beta}\cdot\vec{\rho}_{\nu_\alpha}(t)
\end{equation}
where $\vec{\rho}_{\nu_\alpha}(t)$ is the time evolved state from an initial neutrino flavour $\nu_\alpha$ and 
$\vec{\rho}_{\nu_\beta}$ is the final neutrino flavour $\nu_\beta$ to be detected. Considering that the neutrinos are ultrarelativistic, we have $t=L$ where $L$ is the baseline. 
\subsection{Neutrino Hamiltonian, rotation and CP phase}
The neutrino Hamiltonian in the VMB for a neutrino of energy $E_\nu$ is given by
\begin{equation}\nonumber
H_V(\delta )=\hat{H}_\text{VAC} + \hat{A}
\end{equation}
where $\hat{H}_\text{VAC} =\text{diag}(0,\text{$\Delta $m}_{21}^2/2 E_\nu,\text{$\Delta $m}_{31}^2/2 E_\nu)$, 
$\hat{A}=U^\dagger\text{diag}(A_\text{CC}/2 E_\nu,0,0) U$, with $A_\text{CC}=2E \sqrt{2}G_F n_e $,  where $G_F,n_e$ are the Fermi coupling constant and electron number density, respectively. Here $U$ is the PMNS matrix, parametrized as
\begin{equation}
U=U_{23} U_{\delta }^\dagger U_{13} U_{\delta } U_{12}
\end{equation}
with $U_{\delta}=\text{diag}(1,1,e^{-i\delta})$. Considering that the matrix $U_{23}U_{\delta }^\dagger$ 
conmutes with $\text{diag}(A_\text{CC},0,0)$ and the matrix $U_{\delta }$ conmutes with $U_{12}$ and 
$\text{diag}(0,\text{$\Delta $m}_{21}^2,\text{$\Delta $m}_{31}^2)$ separately, the neutrino Hamiltonian can be re-written as follows 
\begin{equation}
H_V(\delta )=U_{\delta }^\dagger H_V(0)U_{\delta },
\label{factor_delta}
\end{equation} 
where $H_V(0 )$ is given by
\begin{eqnarray}\nonumber
\hspace*{-5cm}
H_V(0) &=& \dfrac{1}{2 E_\nu} \left\{ \left(
\begin{array}{ccc}
0 & 0 & 0 \\
0 & \text{$\Delta $m}_{21}^2 & 0 \\
0 & 0 & \text{$\Delta $m}_{31}^2 \\
\end{array}
\right)\right.\\
&{}&\qquad\quad+\left.U^\dagger_{12}{U^\dagger_{13}}\left(
\begin{array}{ccc}
A & 0 & 0 \\
0 & 0 & 0 \\
0 & 0 & 0 \\
\end{array}
\right){U_{13}}{U_{12}} \right\}.
\end{eqnarray}
The Hamiltonian $H_V(\delta)$ can be diagonalized with the unitary matrix $\textit{U}_T( \delta,\phi_1,\phi_2 ) $ that
rotates the mass eigenstates in the MMB into the mass
eigenstates in the VMB. 
\begin{equation}
H_M (\delta ) = U_T^\dagger (\delta,\phi_1,\phi_2 )H_V(\delta ) \text{\textit{$U$}}_T( \delta,\phi_1,\phi_2 ) 
\end{equation}
Using Eq. (\ref{factor_delta})  and the property that the diagonal matrix $H_M(\delta)$ commutes with the unitary phase operator 
$U^\dagger_\phi (\phi_1,\phi_2)=\text{diag}(1,e^{-i\phi_1},e^{-i\phi_2})$, being $\phi_1$ and $\phi_2$ arbitrary phases, the latter equation 
can be rewritten as follows:
\begin{eqnarray}\nonumber
H_M (\delta ) = &U^\dagger_\phi (\phi_1,\phi_2) U_T^\dagger (\delta,\phi_1,\phi_2 )U_{\delta }^\dagger H_V(0)\\
&\times U_{\delta } U_T (\delta,\phi_1,\phi_2 )U_\phi (\phi_1,\phi_2)
\label{general_UT}
\end{eqnarray}

Now that $H_M(\delta)$ is shown as a unitary transformation of $ H_{V}(0) $, it follows that
the eigenvalues of the former do not depend on $\delta$. We thus write:
\begin{equation}
H_M (\delta ) =H_M (0 ) =U^\dagger_T(0,0,0) H_V(0) \text{\textit{$U$}}_T(0,0,0)  
\label{diagonal_h0}
\end{equation}
Comparing Eqs. (\ref{general_UT}) and (\ref{diagonal_h0}) we get
\begin{equation}
U_T (\delta,\phi_1,\phi_2 )= U^\dagger_\delta  U_T (0,0,0 )  U^\dagger_\phi (\phi_1,\phi_2)\label{unitary_UT}
\end{equation}
This result implies that once we find a matrix $U_T (0,0,0 )$ that diagonalizes $ H_V(0)$, 
a general $U_T (\delta,\phi_1,\phi_2 )$ that diagonalizes $H_V(\delta)$ can be constructed. 

\subsection{Invariant decoherence matrix analysis}
We can apply the results of the preceding section to the problem of finding a decoherence matrix invariant under $U_T$.
As mentioned in Ref. \cite{Carpio18}, the relationship between VMB and MMB decoherence matrices goes as follows:
\begin{equation}
M_D^M(\delta,\phi_1,\phi_2) = P(\delta,\phi_1,\phi_2) M_D^V P^T(\delta,\phi_1,\phi_2).
\label{rotdecmat1}
\end{equation}
where the orthogonal matrix $P$ performs the rotation from the VMB to the MMB and is related to how the $SU(3)$ generators transform 
under $U_T$. The matrix $P(\delta,\phi_1,\phi_2)$ is defined as: 
\begin{equation}
P(\delta,\phi_1,\phi_2) = F(\phi_1,\phi_2) P_0 R(\delta)
\label{ortho_P}
\end{equation}
where the expressions for $P_0=P(0,0,0)$, $R(\delta)$ and $F(\phi_1,\phi_2)$ are given in the appendix. Eq. \eqref{ortho_P}
is the analog of Eq. \eqref{unitary_UT} after performing the $SU(3)$ expansion.
Although we are emphasizing the 
dependence of $M_D^M$ on $\delta$ and the arbitrary phases $\phi_1$ and $\phi_2$ in Eq. \eqref{rotdecmat1}, we must notice 
that this matrix, in general, may depend on the neutrino energy and the matter potential $A$. 

Now we want to prove that, given an invariant matrix $M_D^M=M_D^V$ when $\delta=0$, it is possible to derive invariant matrices $\bar{M}_D^V$ for other $\delta$. This 
situation is very convenient to deal with since the neutrino matter oscillation probability formulae in the presence of quantum decoherence 
are easily found from their corresponding expressions in vacuum.
To prove this, we rewrite Eq. \eqref{rotdecmat1} explicitly as
\begin{eqnarray}\nonumber
M_D^M(\delta,\phi_1,\phi_2)&=&F{(\phi_1,\phi_2)} P_0 R(\delta) M_D^V \\
&{}&\times R^T(\delta) P_0^T F^T{(\phi_1,\phi_2)}
\label{rotdecmat2}
\end{eqnarray}
and set the invariance condition $M_D^M(\delta,\phi_1,\phi_2) = M_D^V$ for $\delta=0$, leading to  
\begin{equation}
M_D^V=F(\phi_1,\phi_2) P_0 M_D^V P_0^T F^T(\phi_1,\phi_2).
\label{invariantdelta0}
\end{equation}
For the case $\delta\neq 0$, the invariance condition for $M_D^M(\delta,\tilde{\phi}_1,\tilde{\phi}_2)=\bar{M}_D^V$ reads
\begin{eqnarray}\nonumber
\bar{M}_D^V&=F(\tilde{\phi}_1,\tilde{\phi}_2) P_0 R(\delta) \bar{M}_D^V \\
&\times R^T(\delta) P_0^T F^T(\tilde{\phi}_1,\tilde{\phi}_2).
\label{invariantdelta}
\end{eqnarray}

Imposing the following relation to the right hand side of the preceding equation
\begin{equation}
\bar{M}_D^V = R^T(\delta) M_D^V R(\delta),
\label{deltaconnection}
\end{equation} 
together with Eq.(\ref{invariantdelta0}), we get:
\begin{eqnarray}\nonumber
M_D^M(\delta,\tilde{\phi}_1,\tilde{\phi}_2) =&F(\tilde{\phi}_1,\tilde{\phi}_2)F^T(\phi_1,\phi_2) M_D^V\\
& \times F(\phi_1,\phi_2)F^T(\tilde{\phi}_1,\tilde{\phi}_2).
\end{eqnarray}  
Thus if we demand $F(\phi_1,\phi_2) F^T(\tilde{\phi}_1,\tilde{\phi}_2)= R(\delta)$ 
the condition $M_D^M(\delta,\tilde{\phi}_1,\tilde{\phi}_2) =\bar{M}_D^V$ is achieved and $\bar{M}_D^V$ is invariant. Such a choice
is certainly possible, for instance $\tilde{\phi}_1=\phi_1,\tilde{\phi}_2=\phi_2+\delta$ (see appendix B).

In Ref. \cite{Carpio18},  it is found that for $\delta=0,\pi$ the invariant matrix $M_D^V$ is given by
\begin{equation}
M_D^V=-\text{Diag}(\Gamma_1,\Gamma_2,\Gamma_1,\Gamma_1,\Gamma_2,\Gamma_1,\Gamma_2,\Gamma_1),
\label{diagvalid0}
\end{equation}
with $\Gamma_1/3\leq\Gamma_2\leq 5\Gamma_1/3$.
Substituting this $M_D^V$  into Eq. \eqref{deltaconnection}, we find that the corresponding invariant $\hat{M}_D^V$ for any 
$\delta$ is described by the following block diagonal matrix: 
\begin{equation}
\bar{M}_D^V=R^T(\delta) M_D^V R(\delta) =-\left(
\begin{array}{cccccc}
\Gamma_1 &   &   &   &   &   \\
& \Gamma_2 &   &   &   &  \\
&   & \Gamma_1 &   &   &   \\
&   &   & Q &  &  \\
&   &   &   & Q &  \\
&   &   &   &   & \Gamma_1 \\
\end{array}
\right)
\end{equation}  
where the submatrix $Q$ is
\begin{equation}
Q=\left(\begin{array}{cc}
\Gamma_1 \cos^2 \delta+\Gamma_2 \sin^2 \delta & (\Gamma_1-\Gamma_2) \cos \delta \sin \delta\\
(\Gamma_1-\Gamma_2) \cos \delta \sin \delta & \Gamma_2 \cos^2 \delta+\Gamma_1 \sin^2 \delta
\end{array}\right)
\label{RotationInvariant}
\end{equation}

From the latter equation we can calculate $\bar{M}_D^V$ for $\delta=\pi/2,3\pi/2$ obtaining:
\begin{eqnarray}\nonumber
\bar{M}_D^V &=&R^T(\delta) M_D^V R(\delta)\\
&=&-\text{Diag}(\Gamma_1,\Gamma_2,\Gamma_1,\Gamma_2,\Gamma_1,\Gamma_2,\Gamma_1,\Gamma_1).
\label{diagvalidpi2}
\end{eqnarray}

\subsection{Adding Hamiltonian terms in the weak coupling limit}
\label{addhamiltonianweak}

At this point it is interesting to make a comparison between the approach given in~\cite{Oliveira18} for defining invariant matrices, and the one presented here and in~\cite{Carpio18}. With the purpose of demanding energy conservation, the following conditions are imposed in \cite{Oliveira18}: 
$[H_{VAC},\hat{\mathcal{V}}_j]=0$ and $[\hat{H}_M,\hat{\mathcal{V}}_j^M]=0$, being that  $\hat{\mathcal{V}}_j$  
and $\hat{\mathcal{V}}_j^M$ are  the dissipator operators written 
in the VMB (for the pure vacuum case) and in the MMB, respectively. To achieve the aforementioned conditions, it is required that 
$\hat{\mathcal{V}}_j$ changes when the matter potential term $\hat{A}$ is added to the vacuum Hamiltonian $H_{VAC}$; all of this in the VMB. 
The new 
dissipator operator is defined by: $\hat{\mathcal{V}}_j^\text{new} = U_T \hat{\mathcal{V}}_j U_T^\dagger $, depending on the matter 
potential and the neutrino energy, in a way that it recovers its vaccum form when it is  rotated to the MMB, this is: $\hat{\mathcal{V}}_j^M  =U_T^\dagger  
\hat{\mathcal{V}}_j^\text{new} U_T = U_T^\dagger U_T \hat{\mathcal{V}}_j U_T^\dagger U_T = \hat{\mathcal{V}}_j$. 
This situation is incongruent with the underlying regime used for describing the neutrinos as an open quantum system, which is the Born-Markov approximation (BM). 
The Born approximation is applied when the environment interacts weakly with the neutrino system \cite{goroni01,benatti2001}, while the Markovianity
assumption implies the use of  the Lindblad form (Eq. \eqref{lindblad}) \cite{lindblad01,alicki}. In 
this evolution, the system is modified by the environment and parameterized by the dissipator operators (i.e. the decoherence matrix), but not 
viceversa. Thereby, the environment is considered unaffected by the system and the addition of the matter potential 
term to the vacuum neutrino Hamiltonian should not change (or in a negligible way)  the dissipator operator $\hat{\mathcal{V}}_j $. 

\subsection{Selected decoherence matrix models}
Our analysis will be limited to diagonal decoherence matrices due to their relative simplicity. While analytical expressions are available for these
matrices in the context of vacuum oscillations \cite{Gago02a}, oscillations in matter can be found through perturbative expansions.
For a generic matrix $M_D^V=-\text{Diag}(\Gamma_1,\Gamma_2,\Gamma_3,\Gamma_4,\Gamma_5,\Gamma_6,\Gamma_7,\Gamma_8)$, we can make an expansion
in the small quantities $\alpha=\Delta m^2_{21}/\Delta m^2_{31},\theta_{13}$ and $\bar\Gamma_i=\Gamma_i L$, assuming the latter is $<$0.1. 
The lowest order contribution to the transition probability $P_{\nu_\mu\nu_e}$ is found to be
\begin{eqnarray}\nonumber
P_{\nu_\mu\nu_e}&=&P_{\nu_\mu\nu_e}^{(0)}+\frac{1}{4}\cos^2\theta_{23}[\bar\Gamma_1+\bar\Gamma_3
+\cos4\theta_{12}(\bar\Gamma_3-\bar\Gamma_1)]\\
&{}&+\frac{1}{12}(1-3\cos2\theta_{23})\bar\Gamma_8
\label{IndependentShift}
\end{eqnarray}
where $P_{\nu_\mu\nu_e}^{(0)}$ is the standard oscillation transition probability. We also notice that only three decoherence parameters of such a matrix dominate the contributions to the probability. Furthermore, these terms do
not depend on the neutrino energy or matter potential, leading to a probability shift across all energies - a property that will help us
qualitatively explain our results.
Using the current values of the standard oscillation(SO) parameters
given in Table~\ref{Table1}, this shift turns out to be positive. 
\subsubsection{Model A}
In light of this, the first one-parameter model we propose for this study (referred from now on as model A) assumes $\Gamma_3=\Gamma_8=0$
\begin{equation}
M_D^V = -\text{Diag}(\Gamma,\Gamma,0,\Gamma/4,\Gamma/4,\Gamma/4,\Gamma/4,0)
\label{OliveiraModel}
\end{equation}
with an approximate expression for the transition probabilities:
\begin{eqnarray}\nonumber
P_{\nu_\mu\nu_e}&=&P_{\nu_\mu\nu_e}^{(0)}+\frac{1}{2}\bar{\Gamma}\cos^2\theta_{23}\sin^22\theta_{12}\\\nonumber
& &-\frac{1}{4}\cos^2\theta_{23}\bar{\Gamma}^2\left[\sin^42\theta_{12}+\sin^24\theta_{12}\frac{\sin^2(A\Delta)}{4A^2\Delta^2}\right]\\\nonumber
& &+\frac{\bar{\Gamma}\theta_{13}\sin2\theta_{23}\sin4\theta_{12}}{4(1-A)A\Delta}\left[\sin(A\Delta)\cos(\delta+A\Delta)\right.\\\nonumber
& &\left.\qquad-A^2\sin\Delta\cos(\delta+\Delta)\right]\\\nonumber
& &-\frac{\alpha\bar{\Gamma}}{2A^2\Delta}\cos2\theta_{12}\cos^2\theta_{23}\sin^22\theta_{12}\\
& &\qquad\times \left(\sin2A\Delta-2A\Delta\right)
\label{AppProb_Analytic}
\end{eqnarray}

In this expression, $\bar{\Gamma}=\Gamma L$,$\Delta = \Delta m^2_{31}L/4E$ and  
$A=A_\text{CC}/\Delta m^2_{31}$. To obtain the antineutrino oscillation probabilities, switch $A\to -A$ and $\delta\to -\delta$.
\subsubsection{Model B}
The second model, from now on named model B, is an invariant matrix based on \eqref{diagvalidpi2}
\begin{equation}
M_D^V=-\text{Diag}(\Gamma,5\Gamma/3,\Gamma,\Gamma,5\Gamma/3,\Gamma,5\Gamma/3,\Gamma).
\label{InvModelMatrix}
\end{equation}
For such a matrix we do not have a transition probability formula akin to \eqref{AppProb_Analytic}, however, we may still use \eqref{IndependentShift} to get a qualitative understanding
of our results referred to this model. 
\section{Simulation and Results}
For DUNE simulations, we will use the optimized beam as described in the CDR \cite{cdr} and make use of the files from Ref. \cite{cdr2}.
We use an exposure of 150 kt MW year exposure each for the 
neutrino and antineutrino channels. Both the neutrino mode (Forward Horn Current - FHC) and antineutrino mode (Reverse Horn Current - RHC)
are used for 3.5 years on each. We are also using both the appearance ($\nu_\mu\to\nu_e$) and disappearance ($\nu_\mu\to\nu_\mu$) channels on neutrino and antineutrino
modes. All NC and CC background rates discussed in detail in Ref. \cite{cdr} Section 3.6.1 are included.
\begin{table}[!h]\begin{center}\begin{tabular}{c|c|c}
Parameter & Value & Error\\\hline
$\theta_{12}$ & $33.62^\circ$ & $0.77^\circ$\\
$\theta_{23}$ & $47.2^\circ$ & $2.9^\circ$\\
$\theta_{13}$ & $8.54^\circ$ & $0.15^\circ$\\
$\dfrac{\Delta m_{21}^2}{10^{-5}\text{eV}^2}$ & $7.40$ & 0.21\\
$\dfrac{\Delta m_{31}^2}{10^{-3}\text{eV}^2}$ & $2.494$ & 0.032\\
$\delta$ & varies & -\\
$\rho$ & $2.97$ g cm$^{-3}$ & -\\
Baseline $L$ & 1300km & -
\end{tabular}
\caption{Relevant parameters assumed for our DUNE analysis.}
\label{Table1}
\end{center}\end{table}

Throughout this section, we use the SO parameters from the global fit based on
data available in January 2018 \cite{Esteban17,NuFit2018} assuming normal ordering (NO) and some modifications. 
The ranges provided are not symmetric about the 
fit values, but we will take them as symmetric by taking the average of the one-sided deviations. Due to the relatively 
unconstrained value of $\delta$, we consider different values for this parameter with special emphasis on $\delta=-\pi/2$, which is rather 
similar to the current best fit value. As such, no priors are assigned to $\delta$. We summarize this information in Table \ref{Table1}.

For all subsequent statistical analyses we will use the GLoBES package \cite{Huber05,Huber07}, while oscillation probabilities are
calculated using NuSQuIDS \cite{Arguelles14}.
\begin{figure}[!t]
\includegraphics[width=0.5\textwidth]{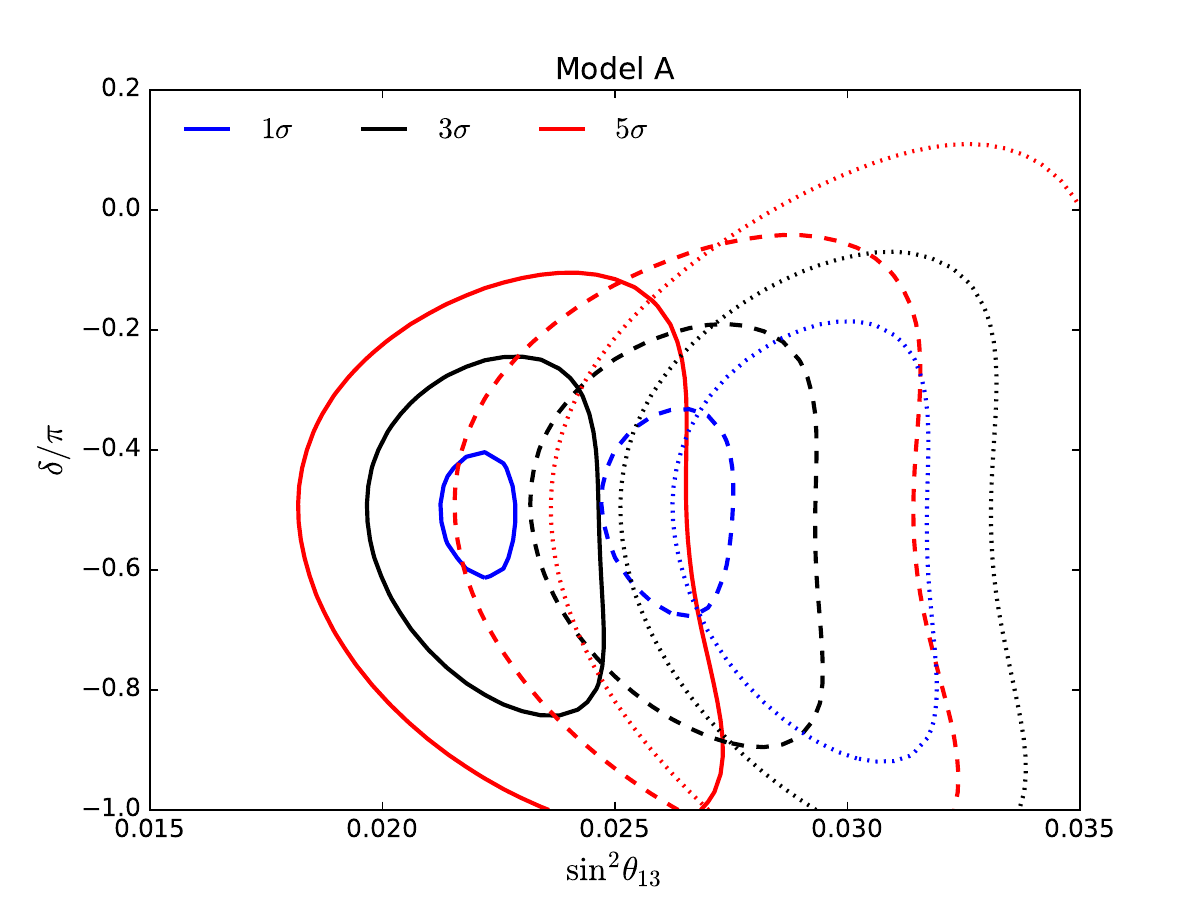}
\includegraphics[width=0.5\textwidth]{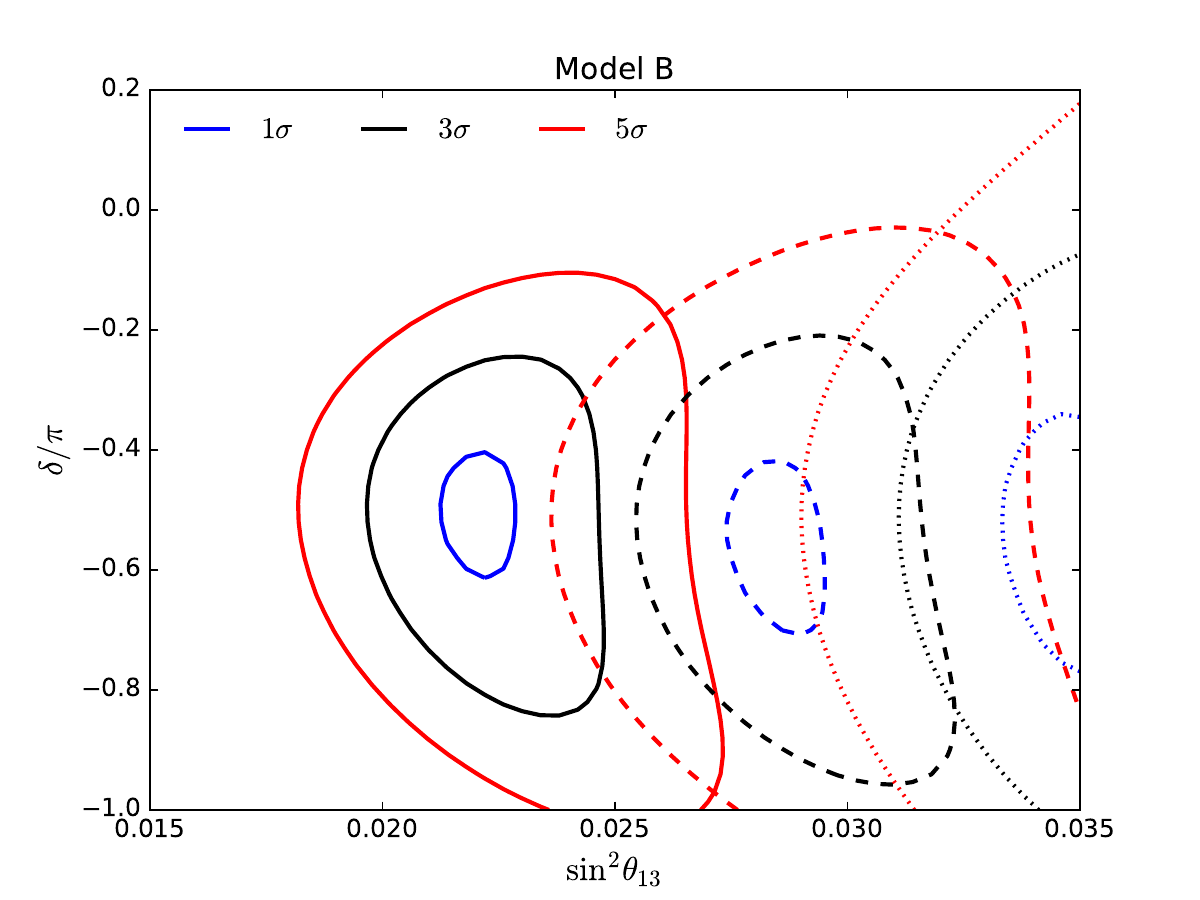}
\caption{Effects of decoherence on standard oscillation fits for $\delta^\text{true}=-\pi/2$. We show the $\Delta\chi^2$ contours
or $\Gamma^\text{true}=0$ GeV, $2\times 10^{-24}$ GeV and $5\times 10^{-24}$ GeV
(solid, dashed and dotted lines respectively).}
\label{DecoVsOsc}
\end{figure}

This study relies on the $\chi^2$ test statistic to compare data generated by a set of ``true'' parameters $\xi^\text{true}$
against a hypothesis that assumes a set of ``test'' values $\xi$. Unless otherwise stated, the parameters $\xi^\text{true}$ will have the 
values in Table \ref{Table1}. The values of
$\delta$ and $\Gamma$, will be provided on a case-by-case basis. In the context of DUNE and GLoBES, $\chi^2$ is defined as
\begin{equation}
\chi^2(\xi,\xi^\text{true})=\sum_i \frac{(N_i(\xi)-N_i(\xi^\text{true}))^2}{N_i(\xi^\text{true})}
\end{equation}
where the $N_i$ is the expected number of events in the i'th energy bin. When including priors, GLoBES modifies $\chi^2$ by adding an 
extra term
\begin{equation}
\chi^2 \to \chi^2 + \sum_j \frac{(\xi_j-\xi_j^\text{true})^2}{\sigma_j^2}
\end{equation}
where the summation in $j$ is over all parameters for which $\sigma_j\neq 0$. In this and all subsequent definitions for $\chi^2$, 
it is implicit that $\Gamma$ and $\gamma$ are used interchangeably depending on the assumed model. 
\begin{figure*}[t]
\vspace{-4pt}
\includegraphics[width=0.8\textwidth]{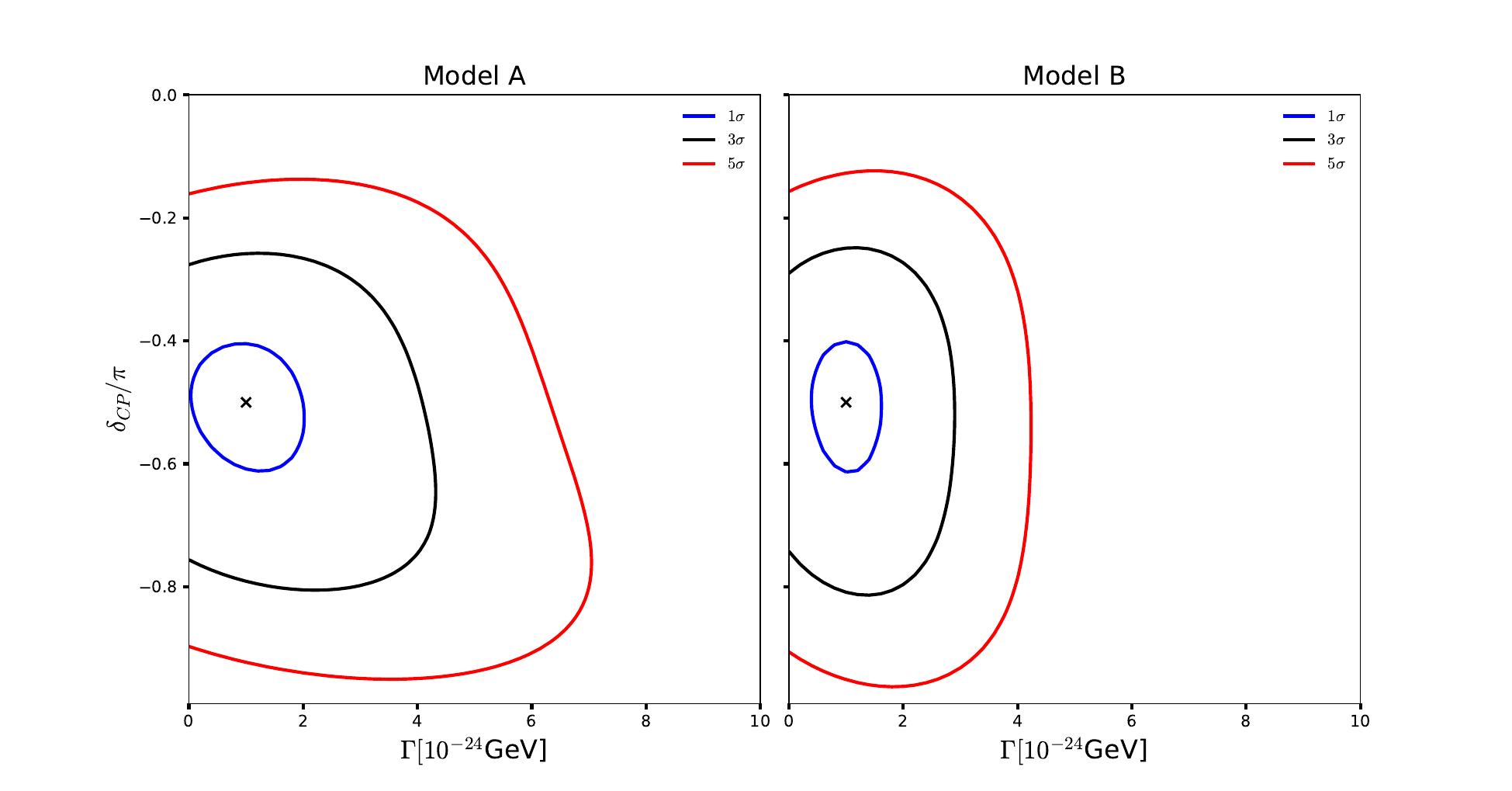}
\caption{Left(right) panel: DUNE's ability to constrain the decoherence parameter for $\Gamma^\text{true}=10^{-24}$ GeV, 
assuming model A(B). The central point $(\delta^\text{true},\Gamma^\text{true})=(-\pi/2,10^{-24}$ GeV) is marked in the graphs.} 
\label{DecoVsDeco}
\end{figure*}

\subsection{Determination of standard oscillation parameters}

Our first test consists on performing SO fits on data that includes decoherence. The data is generated assuming the true parameter values
in Table \ref{Table1}, for $\delta^\text{true}=-\pi/2$ and $\Gamma^\text{true}$ and no priors. The test hypothesis is 
$\Gamma=0$ and we obtain $\chi^2_\text{min}$ by marginalizing over the SO parameters, attaining its minimum for a set of parameters which we 
label as ``fit''. We project the $\chi^2$ function on the $\theta_{13}-\delta$ plane
\begin{eqnarray}\nonumber
\Delta\chi^2 &= &\chi^2(\theta_{13},\delta,\Gamma=0,\delta^\text{true},\Gamma^\text{true})\\
&{}&-\chi^2_\text{min}(\theta_{13}^\text{fit},\delta^\text{fit},\Gamma=0,\delta^\text{true},\Gamma^\text{true})
\end{eqnarray}
where all unmentioned test parameters are fixed to their fit values. 

We plot our results in Fig. \ref{DecoVsOsc}. As we increase the value of $\Gamma^\text{true}$, the SO fit shifts towards 
higher values of $\theta_{13}$ for both models. We can explain this behavior through Eq.\eqref{IndependentShift} where: 
as long as we increase $\Gamma$ the $\nu_\mu\to\nu_e$ transition probability grows in an energy-independent way,
which, if it is fitted under the SO assumption, can be misconstrued as a larger mixing angle $\theta_{13}$. Both plots have similar shapes, where model B
has a more notable shift to higher $\theta_{13}$; a feature that is explained by looking at Eq.
\eqref{IndependentShift}, where the parameters $\Gamma_3,\Gamma_8$ are non-zero and increase the value of the shift.

On the other hand, the general shape of the contour plots remains the same.
We also made the same analysis using the FHC and RHC separately, where the former provided stronger constraints compared to the latter, because of the higher statistics of the former. Similarly, we found that the appearance channels restrict the parameter space better
than the disappearance channels. 

We point out that short baseline neutrino oscillation experiments can measure the value of $\theta_{13}$ quite precisely and will not be
confused with the SO+decoherence scenario, since $\Gamma L$ is much smaller than in DUNE. The precision on $\theta_{13}$ is also expected
to improve by the time DUNE becomes operational. An interesting possibility arises:
if DUNE's measurement of $\theta_{13}$ is incompatible with those obtained from reactor experiments, then decoherence would provide an
explanation for this discrepancy, particularly if DUNE's measured $\theta_{13}$ is larger than the accepted value. 

\subsection{Effects of decoherence on constraining $\delta$}
To test the ability of DUNE to constrain decoherence parameters, we generate data assuming true values for $\delta$ and $\Gamma$ and 
make a $\chi^2$ plot on the $\delta,\Gamma$ plane. Our $\chi^2$ in this case assumes $\delta$ and $\Gamma$ as free parameters and all 
remaining parameters are fixed to their true fit values (there is no marginalization here). Our findings are displayed in 
Fig. \ref{DecoVsDeco}.

If we assume $\Gamma^\text{true}=10^{-24}$ GeV, we find that the data is still compatible with $\Gamma/10^{24}$ GeV = 2.0, 4.3 and 7.0
(1.7, 2.9 and 4.3) for model A (B) at $1,3$ and 5$\sigma$, respectively. We note that model B's limits are more stringent than model
A, a feature explained mostly by the increased contribution from the first order correction to the transition probability (see Eq.
\eqref{IndependentShift}). For this assumed value of $\Gamma^\text{true}$, both models are able to exclude the standard oscillation scenario
at the $1\sigma$ level. 

Even in the presence of decoherence, $\delta$ is well constrained. In \cite{Carpio18} we pointed out that the CP assymmetries for 
$\delta = \pm \pi/2$ were similar for low neutrino energies and it was possible to confuse these two values of $\delta$ in the presence
of decoherence. We also mentioned that after taking the whole of DUNE's energy range into account there was no similarity remaining. 
The aforementioned point is consistent with our current results, where we found that our plots have no secondary contours appearing in the
vicinity of $\delta = \pi/2$.

\subsection{Constraining the decoherence parameter}
\begin{figure}[!t]
\vspace{-4pt}
\includegraphics[width=0.5\textwidth]{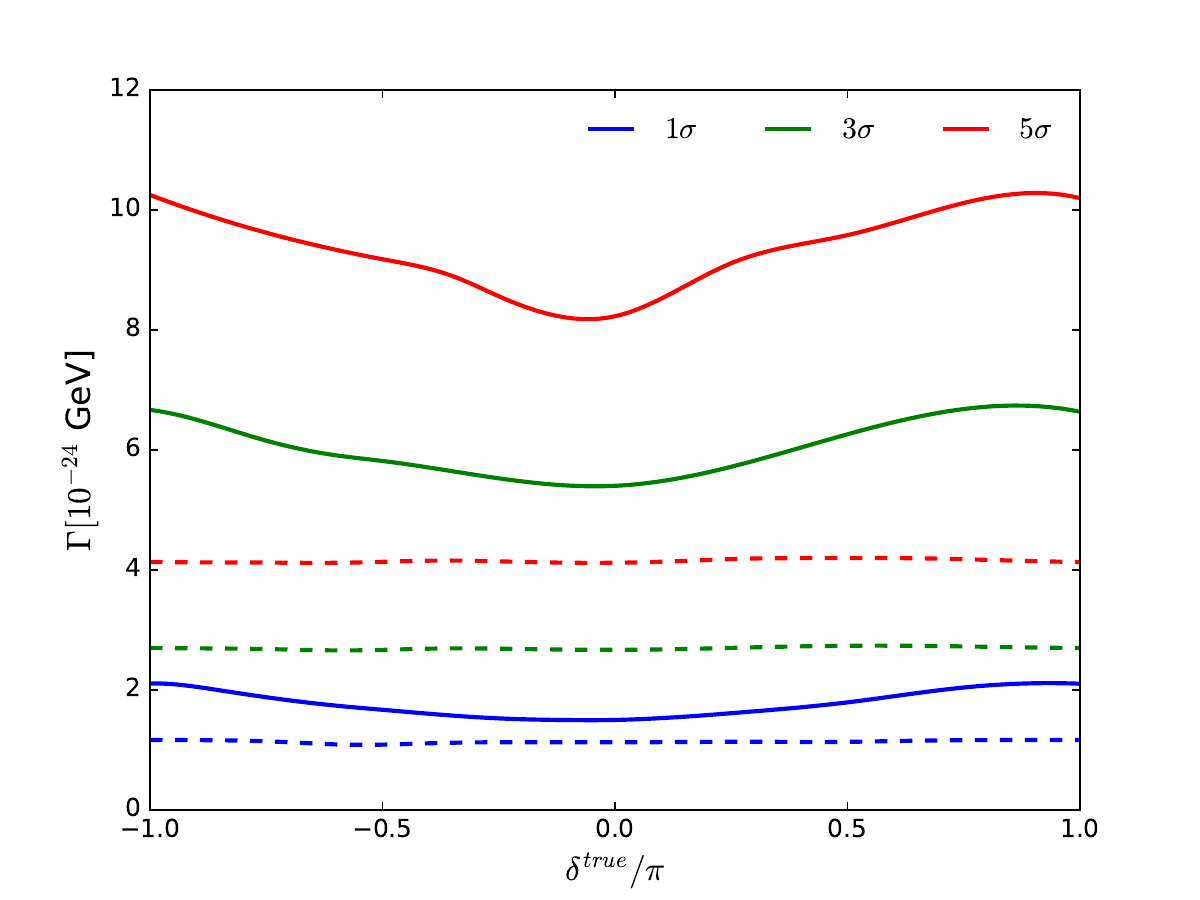}
\caption{Sensitivity
to the decoherence parameter $\Gamma$, as a function of $\delta^\text{true}$, for different confidence levels. Solid lines correspond to
model A and dashed lines correspond to model B.}
\label{DecoSensitivity}
\end{figure}

For this study the simulated data assumes $\Gamma^\text{true}=0$ and a fit is performed for a given test parameter $\Gamma$. Our test 
statistic is
\begin{equation}
\chi^2(\Gamma) = \chi^2(\Gamma,\delta^\text{true},\Gamma^\text{true}=0)
\end{equation}
where we marginalize over all standard oscillation parameters. We show the results in Fig. \ref{DecoSensitivity}. The sensitivity to 
$\Gamma/(10^{-24}\text{GeV})$ for the model A is in the range 1.5-2.1, 5.4-6.7 and 8.2-10.3 at the $1\sigma,3\sigma$ and $5\sigma$ levels. 
On the other hand, the mode of the invariant matrix, model B, has a sensitivity almost invariant in $\delta$ and it is stronger than one predicted for model A, as we have already seen it in Fig. \ref{DecoVsDeco}. This is a result of the increased contribution from
the first order decoherence term for model B, which does not depend on the CP phase; the effect of other $\delta$-dependent terms in a
perturbative expansion are suppressed, such as the $\Gamma\theta_{13}$ term present in Eq. \eqref{AppProb_Analytic}.

For either model, DUNE's sensitivity to $\Gamma$ is superior to the previously reported KamLAND's 95\% C.L. 
sensitivity of $6.8\times 10^{-22}$GeV  \cite{Oliveira16} and the 90\% C.L. sensitivity of $1.2\times 10^{-23}$ GeV \cite{Oliveira18}.
Our current results are also of the same order of magnitude as those reported in a recent IceCube study \cite{Coloma18}.
We do remark that the decoherence models used in these previous studies are slightly different than ours and act as benchmarks. 

\subsection{Mass ordering sensitivity}

Mass ordering sensitivity is obtained by comparing generated data from an NO assumption with decoherence against an IO hypothesis 
without decoherence. When generating data, a value of $\Gamma^\text{true}$ is assumed and the test statistic is marginalized in all 
oscillation parameters while keeping $\Gamma=0$ fixed, meaning that
\begin{equation}
\chi^2_\text{MO} = \chi^2(\Delta m_{31}^2<0,\Gamma=0,\Delta m_{31}^{2\text{true}}>0,\delta^\text{true},\Gamma^\text{true})
\end{equation}

\begin{figure*}
\vspace{-4pt}
\centerline{
\includegraphics[width=\textwidth]{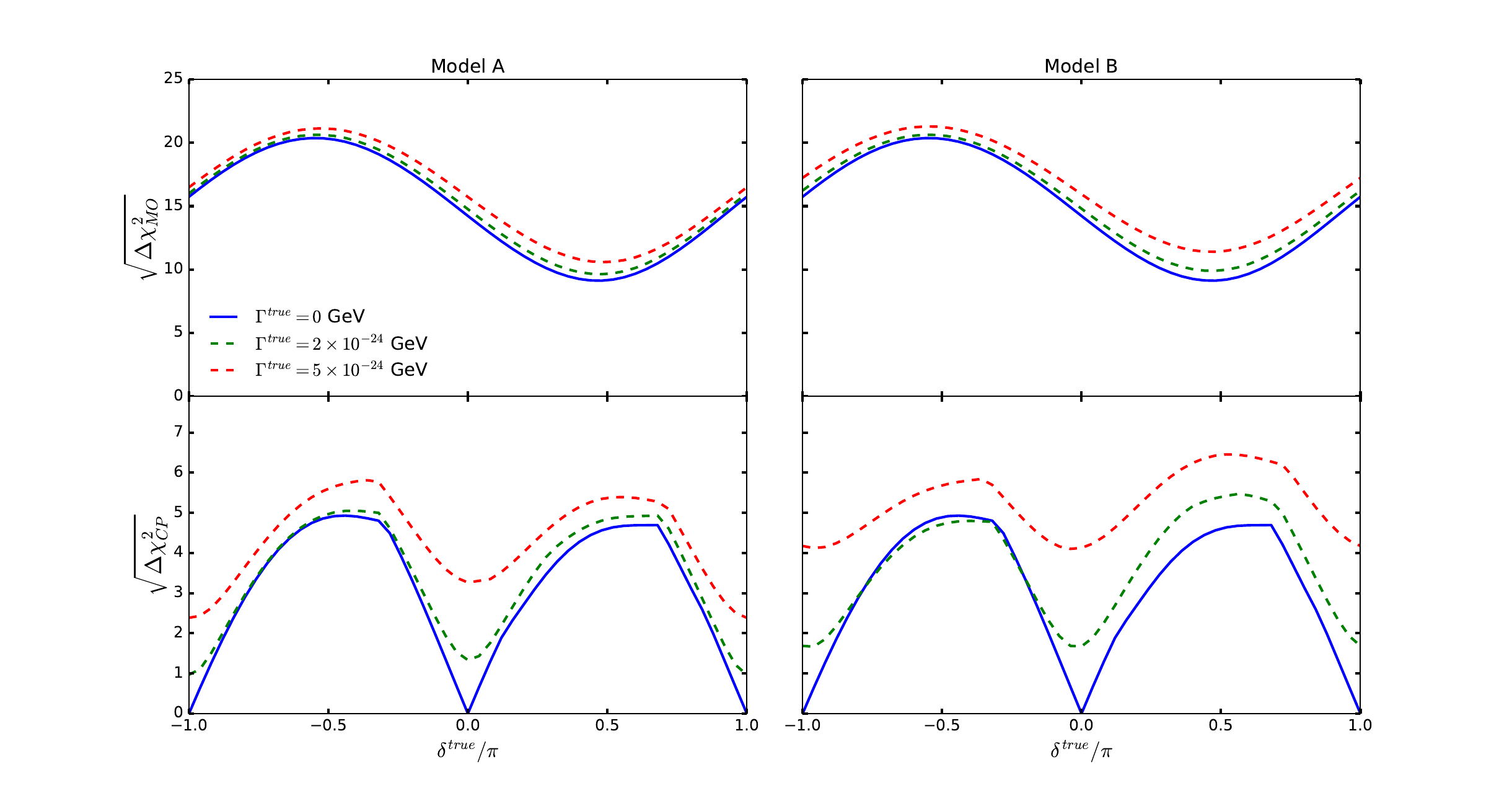}
}
\caption{Mass Hierarchy and CP violation sensitivity for our decoherence matrix models.}
\label{MHCPSensitivity}
\end{figure*}

The sensitivities are presented in Fig. \ref{MHCPSensitivity} for different  $\Gamma^\text{true}$. We see that, regardless of 
$\delta^\text{true}$ and our assumed decoherence model, the sensitivity is well above $5\sigma$ for our chosen values of the 
decoherence parameter and in fact improves it.
To explain this feature, we point out that the IO fit yielded similar values for the SO parameters for all $\delta^\text{true}$, 
with the fit value $\delta^\text{fit}$ in the vicinity of $\delta=-\pi/2$. This occured for all the $\Gamma^\text{true}$ that we analyzed.
When we studied the IO fit, we noticed that the probability in the $\nu_\mu\to\nu_e$ channel is lower, and well separated, than that 
corresponding to the NO data in the energy range
2 GeV-4 GeV. Being that, as long as $\Gamma^\text{true}$ increases, the NO transition probabilities do as well. In return, the resulting IO fit transition
probabilities separate more and more from the corresponding NO assumptions from the data. This separation leads to the observed increase
in sensitivity. The impact of $\Gamma$ on the sensitivity is strongest as we approach $\delta^\text{true}=\pi/2$. 

\subsection{CP violation sensitivity}
To obtain DUNE's sensitivity to CP violation, we adopt a definition similar to the one presented in~\cite{cdr}
\begin{eqnarray}\nonumber
\chi^2_{CP} = \text{min}[\chi^2(\delta=0,\Gamma=0,\delta^\text{true},\Gamma^\text{true}),\\
\chi^2(\delta=\pi,\Gamma=0,\delta^\text{true},\Gamma^\text{true})]
\end{eqnarray}
and we marginalize over all unmentioned test parameters. We checked that, after marginalization, the oscillation parameters remain
approximately
the same, with the exception of $\theta_{13}$ that settles at higher values to account for the increased event rates in the appearance
channels due to decoherence. The IO alternative was never preferred during the minimization
procedure of $\chi^2$. This observation is easily explained by Fig. \ref{MHCPSensitivity} because DUNE can distinguish between
the hierarchies with a large significance. 

The data generated using NO has an additional fake CP violation contribution coming from the matter potential present in the term proportional to $\Gamma\theta_{13}$, which is mistaken in an SO fit
as CP violation. For this reason, even when $\delta^\text{true}=0,\pi$, decoherence may cause us to reject the null hypothesis 
(no CP violation). Due to this decoherence induced CP violation,  decoherence lets us reject the null hypothesis at a higher confidence level
throughout all values of $\delta$. 

\section{SUMMARY AND CONCLUSIONS}
We have tackled diverse aspects of the decoherence phenomenon in the context of neutrino oscillations. From the theoretical side, we 
derived a general expression for invariant decoherence matrices that are diagonal (under rotations from VMB to MMB). The advantage of this 
type of matrix is its simple implementation into the matter oscillation probabilities formula as clarified in
\cite{Carpio18}. On the other hand, we have probed an invariant and non-invariant decoherence matrix in the context of DUNE.
We have achieved a sensitivity for the decoherence parameter of $\mathcal{O}(10^{-24}$ GeV) at 3$\sigma$ and for 
$\delta =-\pi/2$, in the best case scenario (the invariant matrix model). A very interesting result is the presence of an interplay between
$\Gamma$ and $\delta$, predicted at the theoretical level in \cite{Carpio18}. Additionally we found that, if decoherence plus 
standard oscillation is embodied within the data, the pure standard oscillation fit tends to select higher values of $\sin^2\theta_{13}$ 
compared to the pure oscillation case. Finally, we have also seen that the decoherence phenomenon mildly affects the DUNE sensitivity on 
the mass hierarchy, while the CP sensitivity is clearly improved but with the disadvantage that even a true lack of CP violation 
($\delta^\text{true}=0,\pi$) can be confused with CP violation.      
    
\section{ACKNOWLEDGEMENTS}
The authors acknowledge funding by the {\it Direcci\'on de Gesti\'on de la Investigaci\'on} at PUCP, through grant  DGI-2017-3-0019. They would also like to thank F. de Zela, J.~Jones-P\'erez and C.~Arg\"uelles for useful discussions and a very careful reading of the manuscript. 

\appendix

\section{Rotation matrices}
We use the following forms of the unitary operators $U_{ij}$:
\begin{eqnarray}\nonumber
U_{23} =\left(
\begin{array}{ccc}
1 & 0 & 0 \\
0 & \cos \theta_{23} & \sin \theta_{23} \\
0 & -\sin \theta_{23} & \cos \theta_{23}\\
\end{array}
\right)\\\nonumber
U_{13}  
=
\left(
\begin{array}{ccc}
\cos \theta_{13}& 0 & \sin \theta_{13}\\
0 & 1 & 0 \\
-\sin \theta_{13} & 0 & \cos \theta_{13}\\
\end{array}
\right)\\
U_{12}=\left(
\begin{array}{ccc}
\cos\theta_{12}& \sin \theta_{12} & 0 \\
-\sin \theta_{12} & \cos\theta_{12} & 0 \\
0 & 0 & 1 \\
\end{array}
\right)
\end{eqnarray}
\section{Definitions of $P$, $R(\delta)$ and $\hat{R}(\phi)$}
We define the two-dimensional rotation matrix $s(\theta)$ by
\begin{equation}
s(\theta)=\left(\begin{array}{cc}
\cos\theta & \sin\theta\\
-\sin\theta & \cos\theta
\end{array}\right)
\end{equation}
The elements of $P$ are defined as follows:
\begin{equation}
P_{ij}(\delta,\phi_1,\phi_2)= 2\text{Tr}\left[U_T^\dagger(\delta,\phi_1,\phi_2) \tau_j U_T (\delta,\phi_1,\phi_2) \tau_i\right]
\end{equation}
where $\tau_k = \lambda_k/2$ and $\lambda_k$ are the Gell-Mann matrices.
Now replacing $U_T (\delta,\phi_1,\phi_2 )= U^\dagger_\delta  U_T (0,0,0 )  U^\dagger_\phi (\phi_1,\phi_2)$ in the above equation, we obtain: 
\begin{equation}
P(\delta,\phi_1,\phi_2) =F(\phi_1,\phi_2) P(0,0,0) R(\delta)
\end{equation}
where $F,R$ are written in block-diagonal form
\begin{equation}
R(\delta )=\left(
\begin{array}{cccc}
\mathbb{I}_{3x3} &   &   &  \\
& s(\delta) & & \\
& & s(\delta) & \\
& & & 1
\end{array}
\right)
\end{equation}
\begin{equation}
F(\phi_1,\phi_2)=\left(\begin{array}{ccccc}
s(-\phi_1) & & & &\\
&1 &&&\\
&&s(-\phi_2) &&\\
&&& s(\phi_1-\phi_2) &\\
&&&& 1
\end{array}
\right)
\end{equation}


\begin{thebibliography}{}
\bibitem{Fukuda01}
S. Fukuda et al. (Super-Kamiokande Collaboration), Phys. Rev. Lett. \textbf{86}, 5651 (2001).
\bibitem{Ahmad02}
Q.R. Ahmad et al. (SNO Collaboration), Phys. Rev. Lett. \textbf{89}, 011302 (2002).
\bibitem{Fukuda98}
Y. Fukuda et al. (Super-Kamiokande Collaboration), Phys. Rev. Lett. \textbf{81}, 1562 (1998).
\bibitem{Kajita16}
T. Kajita et al. (Super-Kamiokande Collaboration), Nucl. Phys. B \textbf{908}, 14 (2016).
\bibitem{Araki05}
T. Araki et al. (KamLAND Collaboration), Phys. Rev. Lett. \textbf{94}, 081801 (2005).
\bibitem{An12}
F.P. An et al. (Daya Bay Collaboration), Phys. Rev. Lett. \textbf{108}, 171803 (2012).
\bibitem{Adamson14}
P. Adamson et al. (MINOS Collaboration), Phys. Rev. D \textbf{77}, 072002 (2008).
\bibitem{Berryman15}
J. M. Berryman, A. de Gouv\^ea, D. Hern´andez and R. L. N. Oliveira, Phys. Lett. B \textbf{742}, 74 (2015).
\bibitem{Frieman88}
J. A. Frieman, H. E. Haber and K. Freese, Phys. Lett. B \textbf{200}, 115 (1988).
\bibitem{Raghavan88}
R. Raghavan, X.-G. He and S. Pakvasa, Phys. Rev. D \textbf{38}, 1317 (1988).
\bibitem{Berezhiani92a}
Z. Berezhiani, G. Fiorentini, M. Moretti and A. Rossi, Z. Phys. C \textbf{54}, 581 (1992).
\bibitem{Berezhiani92b}
Z. Berezhiani, G. Fiorentini, A. Rossi and M. Moretti, JETP Lett. \textbf{55}, 151 (1992).
\bibitem{Berezhiani93}
Z. G. Berezhiani and A. Rossi. 	arXiv:hep-ph/9306278.
\bibitem{Barger99}
 V. D. Barger et al., Phys. Lett. B \textbf{462}, 109 (1999).
\bibitem{Beacom02}
J. F. Beacom and N. F. Bell, Phys. Rev. D \textbf{65}, 113009 (2002).
\bibitem{Joshipura02}
A. S. Joshipura, E. Masso and S. Mohanty, Phys. Rev. D \textbf{66}, 113008 (2002).
\bibitem{Bandyopadhyay03}
A. Bandyopadhyay, S. Choubey and S. Goswami, Phys. Lett. B \textbf{555}, 33 (2003).
\bibitem{Ando04}
S. Ando, Phys. Rev. D \textbf{70}, 033004 (2004).
\bibitem{Fogli04}
G. Fogli, E. Lisi, A. Mirizzi and D. Montanino, Phys. Rev. D \textbf{70}, 013001 (2004).
\bibitem{Palomares05}
S. Palomares-Ruiz, S. Pascoli and T. Schwetz, JHEP \textbf{0509}, 048 (2005).
\bibitem{Gonzalez08}
M. C. Gonzalez-Garcia and M. Maltoni, Phys. Lett. B \textbf{663}, 405 (2008).
\bibitem{Maltoni08}
M. Maltoni and W. Winter, JHEP0807, 064 (2008).
\bibitem{Baerwald12}
P. Baerwald, M. Bustamante and W. Winter, JCAP \textbf{1210}, 020 (2012).
\bibitem{Meloni07}
D. Meloni and T. Ohlsson, Phys. Rev. D \textbf{75}, 125017 (2007).
\bibitem{Das11}
C.R. Das and J. Pulido, Phys. Rev. D \textbf{83}, 053009 (2011).
\bibitem{Dorame13}
L. Dorame, O.G. Miranda and J.W.F. Valle, Front. in Phys. \textbf{1}, 25 (2013).
\bibitem{Gomes15}
R.A. Gomes, A.L.G. Gomes and O.L.G. Peres, Phys. Lett. B \textbf{740}, 345 (2015).
\bibitem{Picoreti16}
R. Picoreti, M.M. Guzzo, P.C. de Holanda and R.L.N. Oliveira, Phys. Lett B \textbf{761}, 70 (2016).
\bibitem{Abrahao15}
T. Abrah\~ao, H. Minakata and A.A. Quiroga, JHEP \textbf{11}, 001 (2015).
\bibitem{Bustamante17}
M. Bustamante, J.F. Beacom and K. Murase, Phys. Rev. D \textbf{95}, 063013 (2017).
\bibitem{Gago17}
A.M. Gago, R.A. Gomes, A.L.G. Gomes, J. Jones-Perez and O.L.G. Peres (2017). arXiv:1705.03074.
\bibitem{Coloma17}
P. Coloma and O.L.G. Peres (2017). arXiv: 1705.03599
\bibitem{Ascencio-Sosa:2018lbk} 
  M.~V.~Ascencio-Sosa, A.~M.~Calatayud-Cadenillas, A.~M.~Gago and J.~Jones-Pérez,
  Eur.\ Phys.\ J.\ C {\bf 78}, no. 10, 809 (2018)
\bibitem{Gonzalez99}
M. C. Gonzales-Garcia, M. M. Guzzo, P. Krastev and H. Nunokawa, Phys. Rev. Lett. \textbf{82}, 3202 (1999).
\bibitem{Bergmann00}
S. Bergmann, M. M. Guzzo, P. C. de Holanda, P. Krastev and H. Nunokawa, Phys. Rev. D \textbf{62}, 073001 (2000).
\bibitem{Guzzo04}
M. M. Guzzo, P. C. de Holanda and O. L. G. Peres, Phys. Lett. B \textbf{591}, 1 (2004).
\bibitem{Gago01b}
A. M. Gago, M. M. Guzzo, H. Nunokawa, W. J. C. Teves and R. Zukanovich Funchal, Phys Rev D \textbf{64}, 073003 (2001).
\bibitem{Gago02b}
A. M. Gago et al., Phys. Rev. D \textbf{65}, 073012 (2002).
\bibitem{Ohlsson13}
T. Ohlsson, Rept. Prog. Phys. \textbf{76}, 044201 (2013).
\bibitem{Esmaili13}
A. Esmaili and A. Y. Smirnov, JHEP \textbf{1306}, 026 (2013).
\bibitem{Colladay:1996iz} 
  D.~Colladay and V.~A.~Kostelecky,
  Phys.\ Rev.\ D {\bf 55}, 6760 (1997),
\bibitem{Coleman:1998ti} 
  S.~R.~Coleman and S.~L.~Glashow,
  Phys.\ Rev.\ D {\bf 59}, 116008 (1999).
\bibitem{Coleman:1997xq} 
  S.~R.~Coleman and S.~L.~Glashow,
  Phys.\ Lett.\ B {\bf 405}, 249 (1997)
\bibitem{Colladay:1998fq}
  D.~Colladay and V.~A.~Kostelecky,
  Phys.\ Rev.\ D {\bf 58}, 116002(1998).
 
\bibitem{Adamson:2010rn}
  P.~Adamson {\it et al.} [MINOS Collaboration],
  Phys.\ Rev.\ Lett.\  {\bf 105}, 151601 (2010)
\bibitem{AguilarArevalo:2011yi} 
  A.~A.~Aguilar-Arevalo {\it et al.} [MiniBooNE Collaboration],
 Phys.\ Lett.\ B {\bf 718}, 1303 (2013).
\bibitem{Li:2014rya} 
  Y.~F.~Li and Z.~h.~Zhao,
  Phys.\ Rev.\ D {\bf 90}, no. 11, 113014 (2014).
\bibitem{Ellis}
J. Ellis, N. E. Mavromatos and D. V. Nanopoulos, Phys. Lett. \textbf{B293}, 37 (1992); Int. J. Mod. Phys. \textbf{A11}, 1489 (1996).
\bibitem{Benattistrings} 
F. Benatti and R. Floreanini, Ann. of Phys. \textbf{273}, 58 (1999)
\bibitem{Hawking1}
S. Hawking, Comm. Math. Phys. \textbf{87}, 395 (1983); Phys. Rev. D \textbf{37}, 904 (1988); Phys. Rev. D \textbf{53}, 3099 (1996); S. Hawking and C. Hunter, Phys. Rev. D \textbf{59}, 044025 (1999).
\bibitem{Benatti00}
F. Benatti and R. Floreanini, JHEP0002, 32 (2000).
\bibitem{Benatti01}
F. Benatti and R. Floreanini, Phys. Rev. D \textbf{64}, 085015 (2001).
\bibitem{Gago02a}
A.M. Gago, E.M. Santos, W.J.C. Teves and R. Zukanovich Funchal, arXiv:0208166.
\bibitem{Oliveira10}
R.L.N. Oliveira and M.M. Guzzo, Eur. Phys. J. C \textbf{69}, 493 (2010).
\bibitem{Oliveira13}
R.L.N. Oliveira and M.M. Guzzo, Eur. Phys. J. C \textbf{73}, 2434 (2013).
\bibitem{Oliveira16}
R.L.N Oliveira, Eur. Phys. J. C \textbf{76}, 417 (2016).
\bibitem{Lisi00}
E. Lisi, A. Marrone and D. Montanino, Phys. Rev. Lett. \textbf{85}, 1166 (2000).
\bibitem{Farzan08}
Y. Farzan, T. Schwetz and A.Y. Smirnov, JHEP0807, 067 (2008).
\bibitem{Barenboim05}
G. Barenboim and E.N. Mavromatos, JHEP01, 31 (2005).
\bibitem{Bakhti15}
P Bakhti, Y. Farzan and T. Schwetz, JHEP05, 007 (2015).
\bibitem{Gago01a}
A.M. Gago, E.M. Santos, W.J.C. Teves and R. Zukanovich Funchal, Phys. Rev. D \textbf{63}, 073001 (2001).
\bibitem{Morgan06}
D. Morgan, E. Winstanley, J. Brunner and L. F. Thompson, Astropart. Phys. \textbf{25}, 311 (2006).
\bibitem{Fogli07}
G. L. Fogli, E. Lisi, A. Marrone, D. Montanino and A. Palazzo, Phys. Rev. D \textbf{76}, 033006 (2007).
\bibitem{Oliveira14}
R. L. N. Oliveira, M. M. Guzzo, and P. C. de Holanda, Phys. Rev. D \textbf{89}, 053002 (2014).
\bibitem{Balieiro16}
G. Balieiro Gomes, M. M. Guzzo, P. C. de Holanda and R. L. N. Oliveira, Phys. Rev. D \textbf{95}, 113005 (2016).
\bibitem{Guzzo16}
M.M. Guzzo, P.C. de Holanda and R.L.N. Oliveira, Nucl. Phys. B \textbf{908}, 408 (2016).
\bibitem{Carpio18}
J.A. Carpio, E. Massoni and A.M. Gago, Phys. Rev. D \textbf{97}, 115017 (2018).
\bibitem{Oliveira18} 
  G.~Balieiro Gomes, D.~V.~Forero, M.~M.~Guzzo, P.~C.~De Holanda and R.~L.~N.~Oliveira,
  arXiv:1805.09818.
\bibitem{Coloma18}
P. Coloma, J. Lopez-Pavon, I. Martinez-Soler and H. Nunokawa, Eur. Phys. J. C \textbf{78}, 614 (2018).
\bibitem{cdr}
R.~Acciarri {\it et al.} [DUNE Collaboration], arXiv:1512.06148 (2015).
\bibitem{Hernandez16}
S. Hern\'andez, Magister tesis, Pontificia Universidad Cat\'olica del Per\'u, 2016.


\bibitem{goroni01}
V. Gorini, A. Frigerio, M. Verri, A. Kossakowski and E.C.G.Sudarshan, Reports on Mathematical Physics, \textbf{13}, 149, (1978).
High Energy Physics - Phenomenology
\bibitem{benatti2001}
F. Benatti and R. Floreanini,Phys.Rev. D \textbf{64}, 085015 (2001).
\bibitem{lindblad01}
G. Lindblad, Comm. Math. Phys. \textbf{48}, 119 (1976).
\bibitem{alicki}
R. Alicki and K. Lendi, Quantum Dynamical Semigroups and Applications, Lect.
Notes Phys. 286, (Springer-Verlag, Berlin, 1987)
\bibitem{cdr2}
T. Alion {\it et al.}, arXiv:1606.09550 (2016).
\bibitem{Esteban17}
I. Esteban, M.C. Gonzalez-Garcia, M. Maltoni, I. Martinez-Solar and T. Schwertz, J. High Energy Phys. \textbf{01}, 087 (2017).
\bibitem{NuFit2018}
www.nu-fit.org
\bibitem{Huber05}
P. Huber, M. Lindner, W. Winter, Comput. Phys. Commun. \textbf{167}, 195 (2005).
\bibitem{Huber07}
P. Huber, J. Kopp, M. Lindner, \textit{et al.}, Comput. Phys. Commun. \textbf{177}, 432 (2007).
\bibitem{Arguelles14}
C.A. Arguelles Delgado, J. Salvado and C.N. Weaver, arXiv: 1412:3832 (2014) 

\end{thebibliography}
\end{document}